\documentclass[%
 reprint,
preprintnumbers,
nofootinbib,
 amsmath,amssymb,
 aps,
 prl,
]{revtex4-2}

\usepackage{graphicx}
\usepackage{dcolumn}
\usepackage{bm}
\usepackage{hyperref}
\usepackage{xcolor}
\usepackage{natbib}
\usepackage{xspace}
\usepackage{soul}
\usepackage[capitalize]{cleveref}

\usepackage[normalem]{ulem}

\newcommand{\gev}{\;\text{GeV}\xspace}

\newcommand{\BE}{\begin{equation}}
\newcommand{\EE}{\end{equation}}

\newcommand{\CP}{\ensuremath{\mathcal{CP}}\xspace}
\newcommand{\MW}{\ensuremath{M_W}\xspace}
\newcommand{\MZ}{\ensuremath{M_Z}\xspace}
\newcommand{\sltwo}{\ensuremath{\sin^2\theta_\text{eff}^\text{lep}}\xspace}
\newcommand{\GZ}{\ensuremath{\Gamma_Z}\xspace}

\newcommand{\Pdd}{\Phi_1^\dagger\Phi_1}
\newcommand{\Puu}{\Phi_2^\dagger\Phi_2}
\newcommand{\Pdu}{\Phi_1^\dagger\Phi_2}
\newcommand{\Pud}{\Phi_2^\dagger\Phi_1}

\begin{document}

\preprint{DESY-22-065}
\preprint{EFI-22-4}

\title{
New physics effects on the $W$-boson mass\\[.2em]
from a doublet extension of the SM Higgs sector
}
\thanks{This paper is dedicated to the memory of Alberto Sirlin, whose work paved the way for precise predictions of the $W$-boson mass.}

\author{Henning Bahl$^{1}$}
\email{hbahl@uchicago.edu}
\author{Johannes Braathen$^2$}
\email{johannes.braathen@desy.de}
\author{Georg Weiglein$^{2,3}$}
\email{georg.weiglein@desy.de}
\affiliation{$^1$ University of Chicago, Department of Physics, 5720 South Ellis Avenue, Chicago, IL~60637~USA}
\affiliation{$^2$ Deutsches Elektronen-Synchrotron DESY, Notkestr.~85, 22607 Hamburg, Germany\\}
\affiliation{$^3$ II. Institut für Theoretische Physik, Universität Hamburg, Luruper Chaussee 149, 22761 Hamburg, Germany}
\date{\today}

\begin{abstract}
    Recently, the CDF collaboration has reported a new precision measurement of the $W$-boson mass, \MW, showing a large deviation from the value predicted by the Standard Model (SM). In this paper, we analyse possible new physics contributions to \MW from extended Higgs sectors. We focus on the Two-Higgs-Doublet Model (2HDM) as a concrete example. Employing predictions for the electroweak precision observables in the 2HDM at the two-loop level and taking into account further theoretical and experimental constraints, we identify parameter regions of the 2HDM in which the prediction for \MW is close to the new CDF value. We assess the compatibility of these regions with precision measurements of the effective weak mixing angle and the total width of the $Z$ boson.
\end{abstract}

\maketitle


\section{Introduction}

Electroweak precision observables (EWPOs) are crucial for our understanding of the electroweak interactions. Among the EWPOs, the $W$-boson mass \MW, the effective weak mixing angle \sltwo, and the total decay width of the $Z$ boson, \GZ, are especially important. Because of their precise experimental measurements, where the highest accuracy has been reached for \MW, they have a large sensitivity to deviations from their Standard Model (SM) predictions that can be caused by quantum effects of physics beyond the SM (BSM). 

Recently, the CDF collaboration has presented a new measurement of the $W$ boson mass~\cite{CDF:2022hxs} with an unprecedented precision: 
\begin{equation}
M_W = 80.4335 \pm 0.0094 \gev\,. 
\end{equation}
This value deviates by about $7\,\sigma$ from the SM prediction. It is, however, also in tension with some of the previous measurements of \MW and with the average that was obtained in Ref.~\cite{Zyla:2020abc} prior to the announcement of the CDF result. While it will be mandatory to assess the compatibility of the different measurements and to carefully analyse possible sources of systematic effects, the new result from CDF is a strong additional motivation for investigating BSM contributions to the prediction for \MW. In fact, since many years the combined experimental value of \MW has always been consistently above the SM prediction for a Higgs boson mass of about 125~GeV, favouring a non-zero BSM contribution to \MW. The new CDF measurement significantly strengthens the preference for an upward shift compared to the SM prediction arising from BSM effects.

In this paper, we assess the possibility that an extended Higgs sector shifts \MW upwards with respect to the SM. As a concrete example, we focus on the Two-Higgs-Doublet Model (2HDM). The \MW prediction in the 2HDM has been studied at the one- and two-loop level in Refs.~\cite{Frere:1982ma,Bertolini:1985ia,Hollik:1986gg,Hollik:1987fg,Denner:1991ie,Froggatt:1991qw,Chankowski:1999ta,Grimus:2007if,Grimus:2008nb,Lopez-Val:2012uou,Broggio:2014mna,Hessenberger:2016atw,Hessenberger:2018xzo} finding sizeable BSM corrections. In the present paper, we investigate the 2HDM \MW prediction in the context of the CDF measurement,
employing the state-of-the-art two-loop predictions from 
Refs.~\cite{Hessenberger:2016atw,Hessenberger:2018xzo},
as well as other theoretical and experimental constraints.
Since loop effects in the 2HDM affecting \MW will manifest themselves also in the predictions for the effective weak mixing angle and the total width of the $Z$ boson, we also assess the compatibility of those (pseudo-)observables with the experimental data.


\section{Non-standard corrections to \texorpdfstring{$M_W$}{MW} in the 2HDM}

We consider a \CP-conserving 2HDM containing two $SU(2)_L$ doublets $\Phi_1$ and $\Phi_2$ of hypercharge $1/2$. We impose a $\mathbb{Z}_2$ symmetry in the Higgs potential under which the two doublets transform as $\Phi_1\to\Phi_1$, $\Phi_2\to -\Phi_2$, but that is softly broken by an off-diagonal mass term.\footnote{We impose this $\mathbb{Z}_2$ symmetry, as is done commonly in the literature~\cite{Glashow:1976nt,Paschos:1976ay}, in order to avoid tree-level flavour-changing neutral currents that are severely constrained experimentally.} This potential reads
\begin{align}
\label{eq:HiggsPotential}
& V_{\text{2HDM}}(\Phi_1,\Phi_2) =   \\
&= m_{11}^2\,\Pdd + m_{22}^2\,\Puu - m_{12}^2\left(\Pdu + \Pud\right) \nonumber\\
&\hphantom{=} + \frac{1}{2}\lambda_1 (\Pdd)^2 + \frac{1}{2}\lambda_2 (\Puu)^2  + \lambda_3 (\Pdd)(\Puu) \nonumber        \\
&\hphantom{=} + \lambda_4 (\Pdu)(\Pud) + \frac{1}{2}\lambda_5 \left((\Pdu)^2 + (\Pud)^2\right).\nonumber 
\end{align}
All parameters can be assumed to be real, because we focus on the \CP-conserving case. After minimization of the Higgs potential, the Higgs doublets are decomposed as $\Phi_i^T = \left(\phi_i^+, (v_i + \phi_i + i \chi_i)/\sqrt{2}\right)$ with $v_1^2 + v_2^2 \equiv v^2 \simeq 246\gev$ and $v_2/v_1 \equiv \tan\beta$. 

After rotating to the mass eigenstate basis, the Higgs boson spectrum consists of the \CP-even Higgs bosons $h$ and $H$ (obtained by rotating the $\phi_{1,2}$ states by the angle $\alpha$), the \CP-odd $A$ boson and the neutral Goldstone boson $G$ (obtained by rotating the $\chi_{1,2}$ states by the angle $\beta$), as well as the charged Higgs boson $H^\pm$ and the charged Goldstone boson $G^\pm$ (obtained by rotating the $\phi_{1,2}^\pm$ states by the angle $\beta$). We identify
the \CP-even mass eigenstate $h$
with the observed SM-like Higgs boson and work in the so-called alignment limit by enforcing $\alpha = \beta - \pi/2$~\cite{Gunion:2002zf}. The remaining input parameters for our numerical analysis are $m_H$, $m_A$, $m_{H^\pm}$, $\tan\beta$, and $M^2 \equiv m_{12}^2/(\sin\beta\cos\beta)$. Relations between these parameters and the parameters of \cref{eq:HiggsPotential} are listed e.g.\ in Ref.~\cite{Kanemura:2004mg}.

The leading 2HDM corrections to the EWPOs are induced via corrections to the $\rho$ parameter, which is defined as the ratio of the neutral and charged current four-fermion interactions. In the 2HDM, $\rho$ is equal to one at the tree-level. This tree-level value is, however, affected by loop corrections, which are associated with a breakdown of the custodial symmetry. As discussed in detail in Refs.~\cite{Haber:1992py,Pomarol:1993mu,Gerard:2007kn,Grzadkowski:2010dj,Haber:2010bw,Nishi:2011gc,Hessenberger:2016atw}, the custodial symmetry is restored in the scalar sector at the one-loop level if either $m_H = m_{H^\pm}$ or $m_A = m_{H^\pm}$. In the former case where $m_H=m_{H^\pm}$, a restoration of the custodial symmetry in the scalar sector at two loops happens only if the additional constraint of either $\tan\beta=1$ or $m_H^2 = M^2$ is fulfilled.

The non-SM one-loop corrections to the $\rho$ parameter (assuming massless external fermions), $\Delta\rho_{\text{non-SM}}^{(1)}$, in the \CP-conserving 2HDM (and for $\alpha = \beta - \pi/2$) are given by~\cite{Hessenberger:2016atw}
\begin{align} \label{eq:drho1L}
    \Delta\rho_{\text{non-SM}}^{(1)} ={}& \frac{\alpha}{16\pi s_W^2 M_W^2}\bigg\{\frac{m_A^2 m_H^2}{m_A^2 - m_H^2}\ln\frac{m_A^2}{m_H^2} \nonumber\\
    & - \frac{m_A^2 m_{H^\pm}^2}{m_A^2-m_{H^\pm}^2}\ln\frac{m_A^2}{m_{H^\pm}^2} \nonumber\\
    &  - \frac{m_H^2 m_{H^\pm}^2}{m_H^2 - m_{H^\pm}^2}\ln\frac{m_H^2}{m_{H^\pm}^2} +m_{H^\pm}^2\bigg\}\,,
\end{align}
where $s_W$ and $c_W$ are the sine and cosine of the weak mixing angle, respectively, $\alpha \equiv e^2/(4\pi)$, and $e$ is the electric charge. The quantity $\Delta\rho$ enters the prediction for the $W$ boson mass approximately via
\begin{align} \label{eq:dMW_drho}
    \Delta M_W \simeq \frac{1}{2}M_W \frac{c_W^2}{c_W^2 - s_W^2} \Delta\rho\,.
\end{align}
While Eqs.~(\ref{eq:drho1L}) and~(\ref{eq:dMW_drho}) allow for a qualitative understanding of the 2HDM effects on the \MW prediction, a precise higher-order calculation is essential for a comparison with the experimental results. In order to predict \MW (as well as \sltwo and \GZ) we use the code \texttt{THDM\_EWPOS} which is based on Refs.~\cite{Hessenberger:2016atw,Hessenberger:2018xzo}. It incorporates the full one-loop non-SM corrections as well as the leading non-SM two-loop corrections. To be more specific, the two-loop non-SM corrections are calculated in the limit of vanishing electroweak gauge couplings (keeping the ratio of \MW and \MZ constant). Moreover, all quarks and leptons except for the top quark are treated as massless for the non-SM two-loop corrections. For the calculation of the two-loop corrections, all Higgs boson masses are renormalized in the on-shell scheme. The SM corrections are included via the parameterization given in Ref.~\cite{Awramik:2003rn}. They contain the complete one-loop~\cite{Sirlin:1980nh,Marciano:1980pb} and the complete two-loop results~\cite{Djouadi:1987gn,Djouadi:1987di,Kniehl:1989yc,Halzen:1990je,Kniehl:1991gu,Kniehl:1992dx,Halzen:1991ik,Freitas:2000gg,Freitas:2002ja,Awramik:2002wn,Awramik:2003ee,Onishchenko:2002ve,Awramik:2002vu,Bauberger:1996ix,Bauberger:1997ey,Awramik:2006uz}, as well as partial higher-order corrections up to four-loop order~\cite{Avdeev:1994db,Chetyrkin:1995ix,Chetyrkin:1995js,Chetyrkin:1996cf,Faisst:2003px,vanderBij:2000cg,Boughezal:2004ef,Schroder:2005db,Chetyrkin:2006bj,Boughezal:2006xk}.\footnote{See also Refs.~\cite{Weiglein:1998jz,AchimDipl,Chen:2020xzx,Chen:2020xot} for further higher-order contributions involving fermion loops and Ref.~\cite{Degrassi:2014sxa} for a prediction of \MW employing the $\overline{\mbox{MS}}$ renormalisation scheme.}

The remaining theoretical uncertainties of the predictions for \MW, \sltwo, and \GZ arise on the one hand from unknown higher-order contributions. On the other hand, a parametric uncertainty is induced by the experimental errors of the input parameters,  e.g.\ the top-quark mass. Since the discrepancy between the CDF value for \MW and the SM prediction is much larger than those theoretical uncertainties we will not give a detailed account of those uncertainties in the following.


\section{Numerical results}

For our numerical results, we aim at answering the question of whether an $M_W$ value close to the CDF measurement can be obtained in the 2HDM without being excluded by other constraints. A more comprehensive global fit to the electroweak precision data should be carried out in the future once a new world average has been obtained for both the central value of 
$M_W$ and the experimental uncertainty, taking into account the level of compatibility of the individual measurements with each other.

While we expect similar results for all 2HDM types,\footnote{The difference between the 2HDM types appears only in the down-type and lepton Yukawa couplings. Since the two-loop non-SM correction implemented in \texttt{THDM\_EWPOS} uses the approximation of massless down-type quarks and leptons, the choice of the 2HDM type does not affect the EWPO calculation.} we concentrate here for our numerical study on the 2HDM of type I. Regarding our predictions for \MW, we apply various other constraints of both experimental and theoretical nature on the considered parameter space:
\begin{itemize}
  \item vacuum stability~\cite{Barroso:2013awa} and boundedness-from-below~\cite{Branco:2011iw} of the Higgs potential,
  \item NLO perturbative unitarity~\cite{Cacchio:2016qyh, Grinstein:2015rtl},
  \item compatibility of the SM-like scalar with the experimentally discovered Higgs boson using \texttt{HiggsSignals}~\cite{Bechtle:2013xfa,Bechtle:2020uwn},
  \item limits from direct searches for BSM scalars using \texttt{HiggsBounds}~\cite{Bechtle:2008jh,Bechtle:2011sb,Bechtle:2013wla,Bechtle:2020pkv,Bahl:2021yhk},
  \item $b$ physics~\cite{Haller:2018nnx}.\footnote{In practice, the fit results of Ref.~\cite{Haller:2018nnx} are used to obtain $2\sigma$ constraints in the $(m_{H^\pm},\ \tan\beta)$ plane of the 2HDM parameter space. }
\end{itemize}
We use \texttt{ScannerS}~\cite{Muhlleitner:2020wwk} to evaluate all of these constraints apart from the NLO perturbative unitarity constraint, which is evaluated separately. If applicable, we demand that the constraints be fulfilled at the $95\%$ C.L. Taking into account these constraints on the parameter space, we obtain for each parameter point the one- and two-loop predictions for \MW, \sltwo, and \GZ. We note that as \texttt{ScannerS} does not define a renormalisation scheme for the 2HDM mass parameters, we choose to interpret these as on-shell renormalised inputs when used in the two-loop calculations of the EWPOs. 

We perform a random scan of the 2HDM parameter space. While we fix $m_h=125.09\text{ GeV}$ and $\alpha=\beta-\pi/2$, we scan over values of $m_H$ and $m_A$ in the range between $30$ and $1500\gev$, $m_{H^\pm}$ between $150$ and $1500\gev$, $\tan\beta$ between $0.8$ and $50$, and $m_{12}^2$ between $0$ and $4\cdot 10^6\text{ GeV}^2$. All points shown in the Figures pass the theoretical and experimental constraints outlined above. 

\begin{figure*}
    \centering
    \includegraphics[width=.45\textwidth]{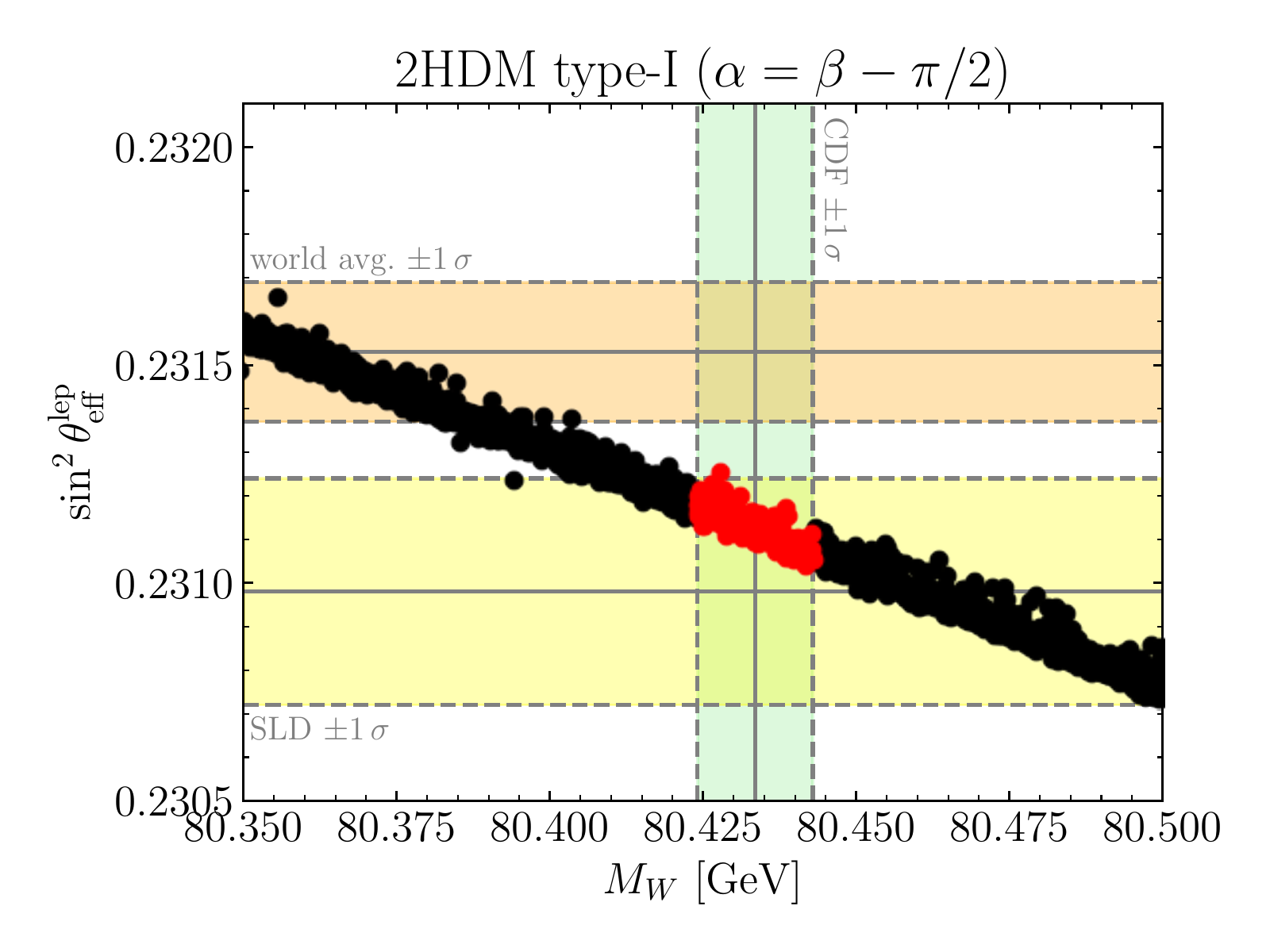} 
    \includegraphics[width=.45\textwidth]{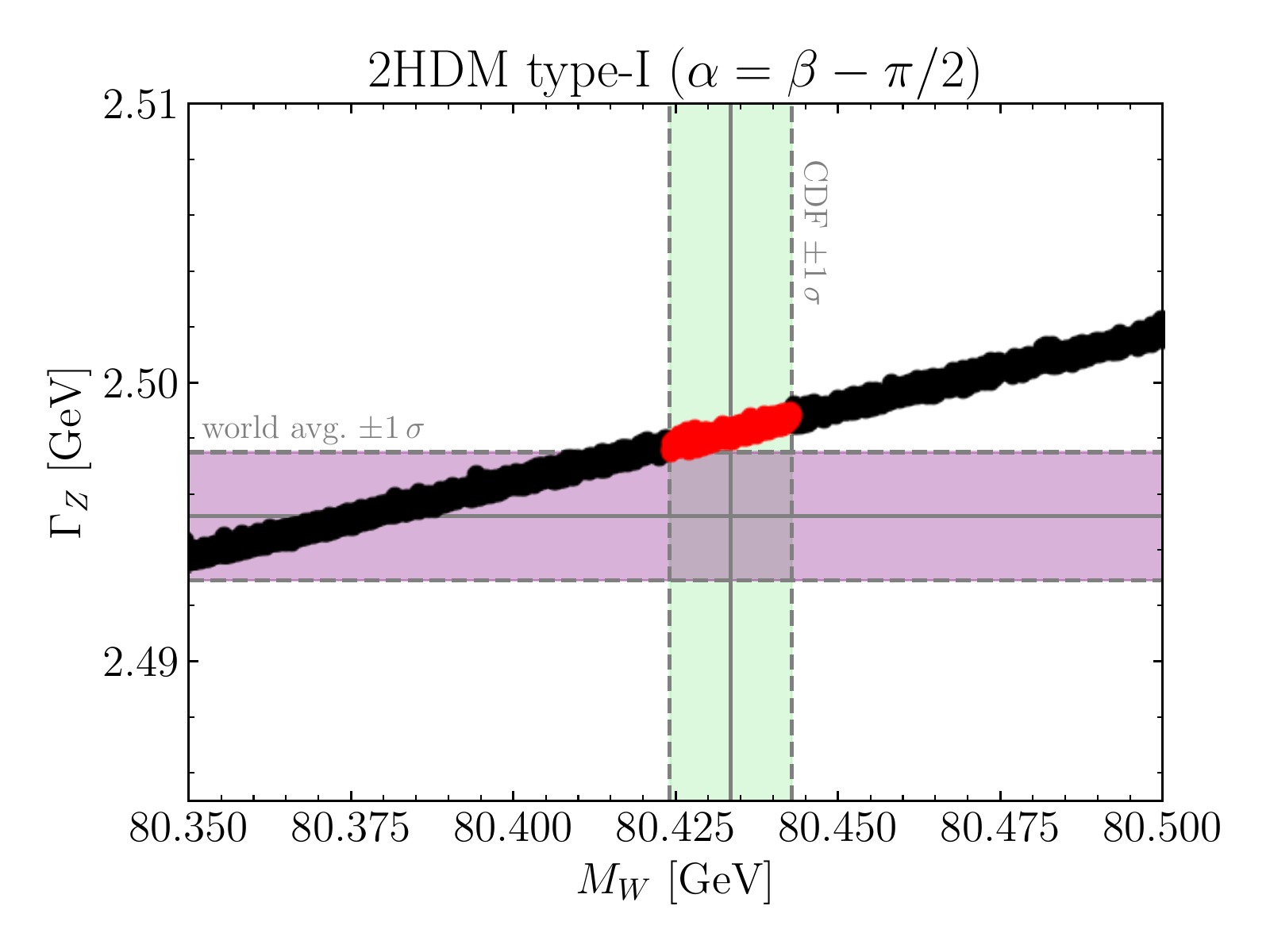} \\
    \includegraphics[width=.45\textwidth]{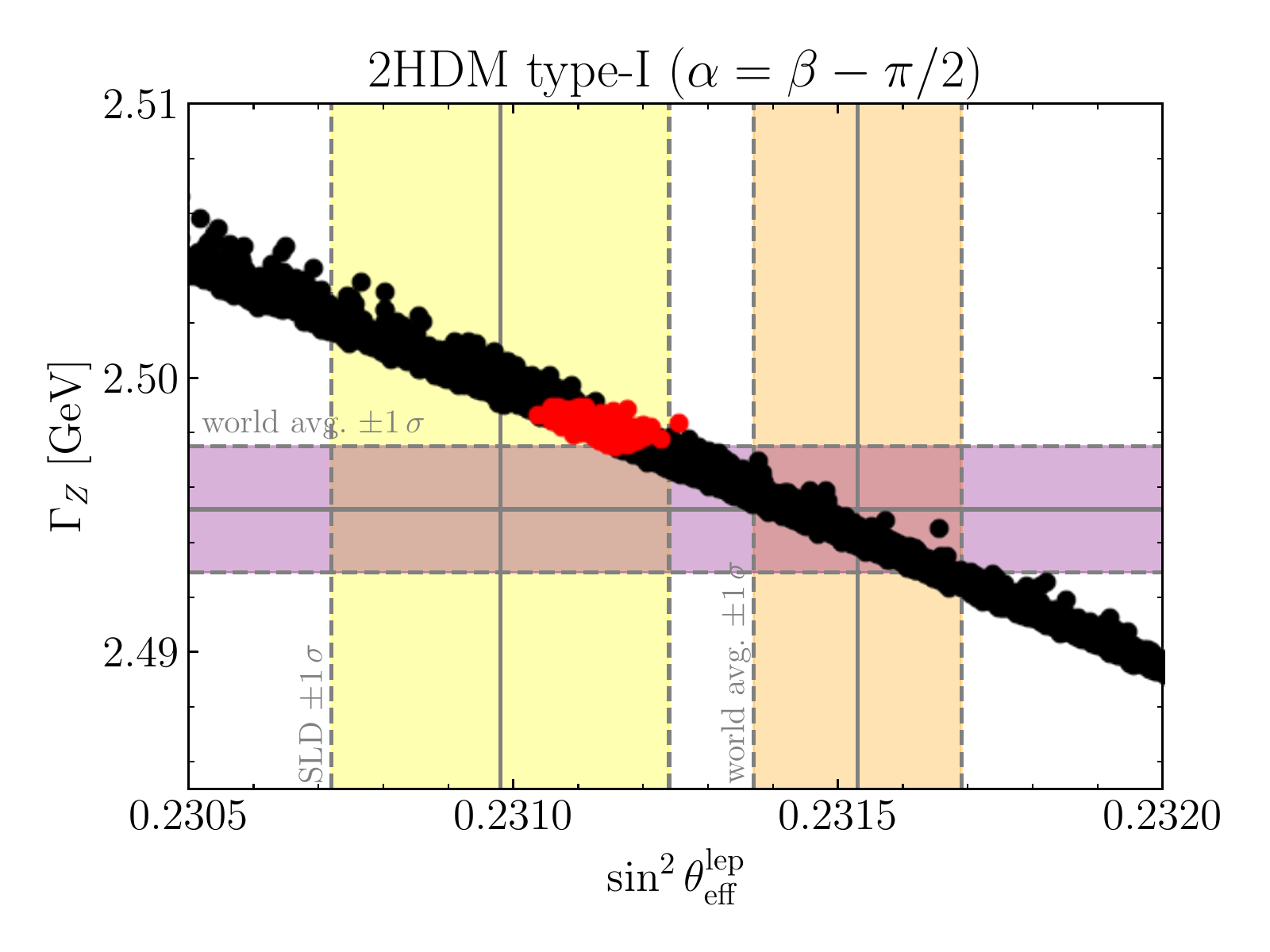}
    \caption{\textit{Upper left:} parameter scan of the type-I 2HDM in the $(\MW, \sltwo)$ plane. The red points are located within the $1\,\sigma$ interval of the recent \MW measurement by the CDF collaboration. \textit{Upper right:} same as upper left panel, but the $(\MW,\GZ)$ plane is shown. \textit{Bottom:} same as upper left panel, but the $(\sltwo, \GZ)$ plane is shown.}
    \label{fig:EWPO}
\end{figure*}

In Fig.~(\ref{fig:EWPO}), we present the scan results for the EWPOs. In the panels, the light green band indicates the \MW value (and the associated $1\,\sigma$ uncertainty) measured recently by the CDF collaboration. Points located within the $1\,\sigma$ interval of the CDF measurement are coloured in red. We also show as a light orange band the world average for \sltwo (and the associated $1\,\sigma$ uncertainty)~\cite{ALEPH:2005ab} that was obtained by averaging over the results of the four LEP collaborations and the SLD collaboration, where the two most precise measurements (based on the forward--backward asymmetry of bottom quarks 
at LEP and the left--right asymmetry at the SLC) showed a discrepancy of more than $3\,\sigma$.
For comparison, we also display the result for \sltwo (and its associated $1\,\sigma$ uncertainty) that is based on the measurement of the left--right asymmetry by the SLD collaboration~\cite{ALEPH:2005ab} as a light yellow band.\footnote{The SLD measurement is the most precise single \sltwo measurement and depends only on leptonic couplings.} In addition, the light purple band shows the world average value for \GZ (and its associated $1\,\sigma$ uncertainty).

In the upper left panel, the results are shown in the $(\MW, \sltwo)$ plane. We observe that an \MW value close to the value measured by the CDF collaboration is very well compatible with the predicted range in the 2HDM (while passing other theoretical and experimental constraints). \MW values within the $1\,\sigma$ interval of the CDF \MW measurement are in mild tension with the world average value for \sltwo but in good agreement with the value measured by the SLD experiment.

In the upper right panel, we show the results in the $(\MW, \GZ)$ plane. We find that the points where \MW is close to the measured CDF value are compatible at the level of $1$--$1.5\,\sigma$ with the world average value for \GZ. This is confirmed in the bottom panel showing the results in the $(\sltwo, \GZ)$ plane. The red points, for which \MW is within $1\,\sigma$ of the CDF value, are at most in mild tension with the results of the other precision measurements at the $Z$~peak.

\begin{figure*}
    \centering
    \includegraphics[width=.45\textwidth]{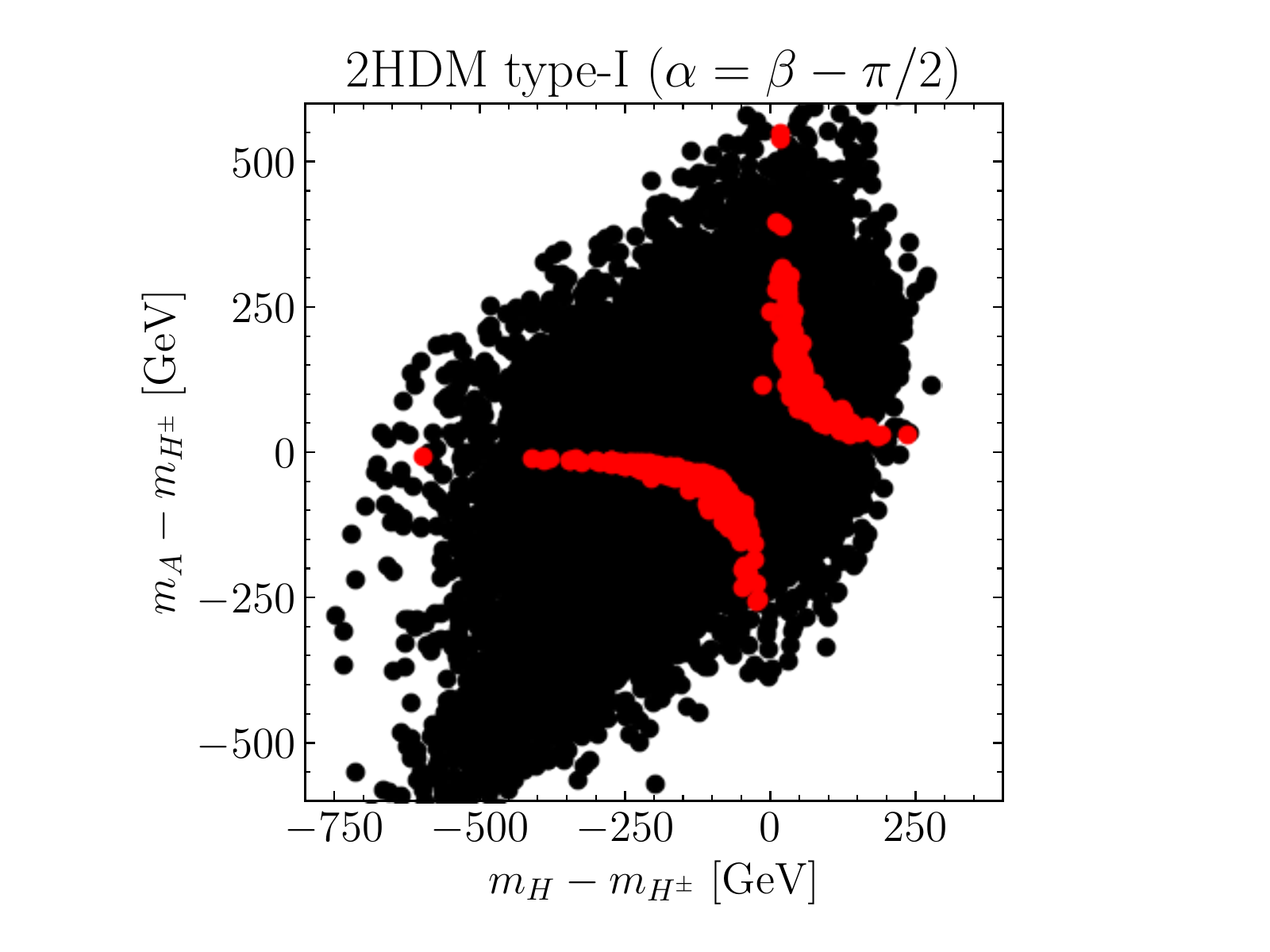} 
    \includegraphics[width=.45\textwidth]{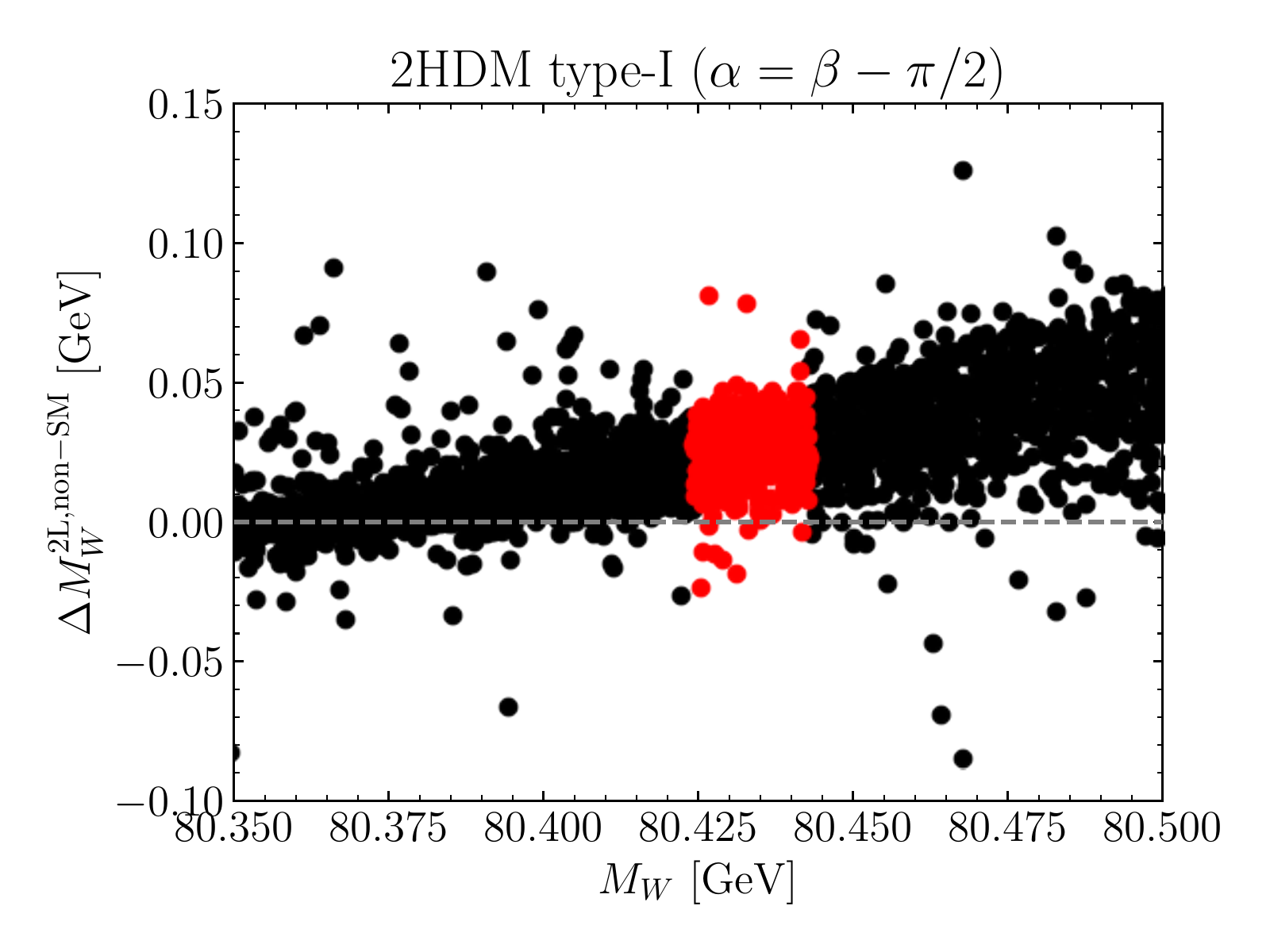}
    \caption{Parameter scan of the type-I 2HDM, where the red points are located within the $1\,\sigma$ interval of the recent \MW measurement by the CDF collaboration. In the left plot, we show the points from the parameter scan in the $(m_H-m_{H^\pm}, m_A-m_{H^\pm})$ plane. In the right plot, we show for the same points the size of the two-loop non-SM corrections to $M_W$ against the total result for $M_W$.}
    \label{fig:cross_2Lvs1L}
\end{figure*}

In the left panel of Fig.~(\ref{fig:cross_2Lvs1L}), we investigate for which mass configurations the CDF value for \MW can be reached. In this panel, the scan results are shown in the $(m_H - m_{H^\pm},\ m_A - m_{H^\pm})$ plane. For $m_H - m_{H^\pm} > 0$ and $m_A - m_{H^\pm} >0$, as well as for $m_H - m_{H^\pm} < 0$ and $m_A - m_{H^\pm} < 0$, the one-loop non-SM correction to the $\rho$ parameter (see Eq.~(\ref{eq:drho1L})) is positive (see also Ref.~\cite{Lu:2022bgw}). Since the CDF value for \MW lies above the SM prediction, it favours this part of the parameter region of the 2HDM giving rise to a sizeable upward shift of \MW, see Eq.~(\ref{eq:dMW_drho}). In contrast, the one-loop non-SM correction is negative for $m_H - m_{H^\pm} > 0$ and $m_A - m_{H^\pm} < 0$ as well as for $m_H - m_{H^\pm} < 0$ and $m_A - m_{H^\pm} > 0$ implying that the CDF \MW value cannot be reached in this part of the parameter space. For reference, we also give in Tab.~(\ref{tab:example_points}) the complete parameter values for two exemplary points in the upper right and lower left part of the left plot of Fig.~(\ref{fig:cross_2Lvs1L}). This Table also shows the \MW values obtained if the non-SM contributions are only evaluated at the one-loop level, finding significantly smaller values. The left plot of Fig.~(\ref{fig:cross_2Lvs1L}) furthermore highlights the fact that the CDF value for $M_W$ cannot be reproduced in scenarios of the 2HDM in which the three BSM scalars are mass-degenerate, since as discussed above the custodial symmetry of the non-SM contribution is restored in the limit where all scalar masses are taken to be equal to each other.

This point moreover implies that there exists an upper bound on the mass of the BSM scalars --- which we estimate 
to be of a few TeV --- in 2HDM scenarios reproducing the CDF value for $M_W$, as absolute mass splittings cannot be maintained for increasingly large masses without violating perturbative unitarity. 
Besides impliying an upper bound on the mass of the BSM scalars, the mass hierarchies favored by the CDF $M_W$ measurement also have interesting implications for the decays of the BSM scalars. Given a large enough mass separation between the charged and the neutral Higgs bosons, $H^\pm\to W^\mp A/H$ or $A/H\to H^\pm W^\mp$ decays, for which the lowest-order couplings are maximised in the alignment limit, have sizeable rates for large parts of the 2HDM parameter space~\cite{Bahl:2021str}. Correspondingly, future experimental searches for these decays could be able to confirm or exclude a 2HDM explanation of large positive shifts of $M_W$ with respect to its SM value.

We further assess the size of the two-loop non-SM corrections in the right panel of Fig.~(\ref{fig:cross_2Lvs1L}) showing the scan results in the $(M_W, \Delta M_W^{\text{2L,non-SM}})$ plane. Here, $\Delta M_W^{\text{2L,non-SM}}$ denotes the difference between \MW evaluated employing the one- and two-loop non-SM correction and \MW evaluated employing only the one-loop non-SM corrections. We observe that large values for \MW are often associated with sizeable positive non-SM two-loop corrections. This shows the importance of a precise evaluation of the non-SM contributions to \MW taking into account corrections beyond the one-loop level.

\begin{figure*}
    \centering
    \includegraphics[width=.45\textwidth]{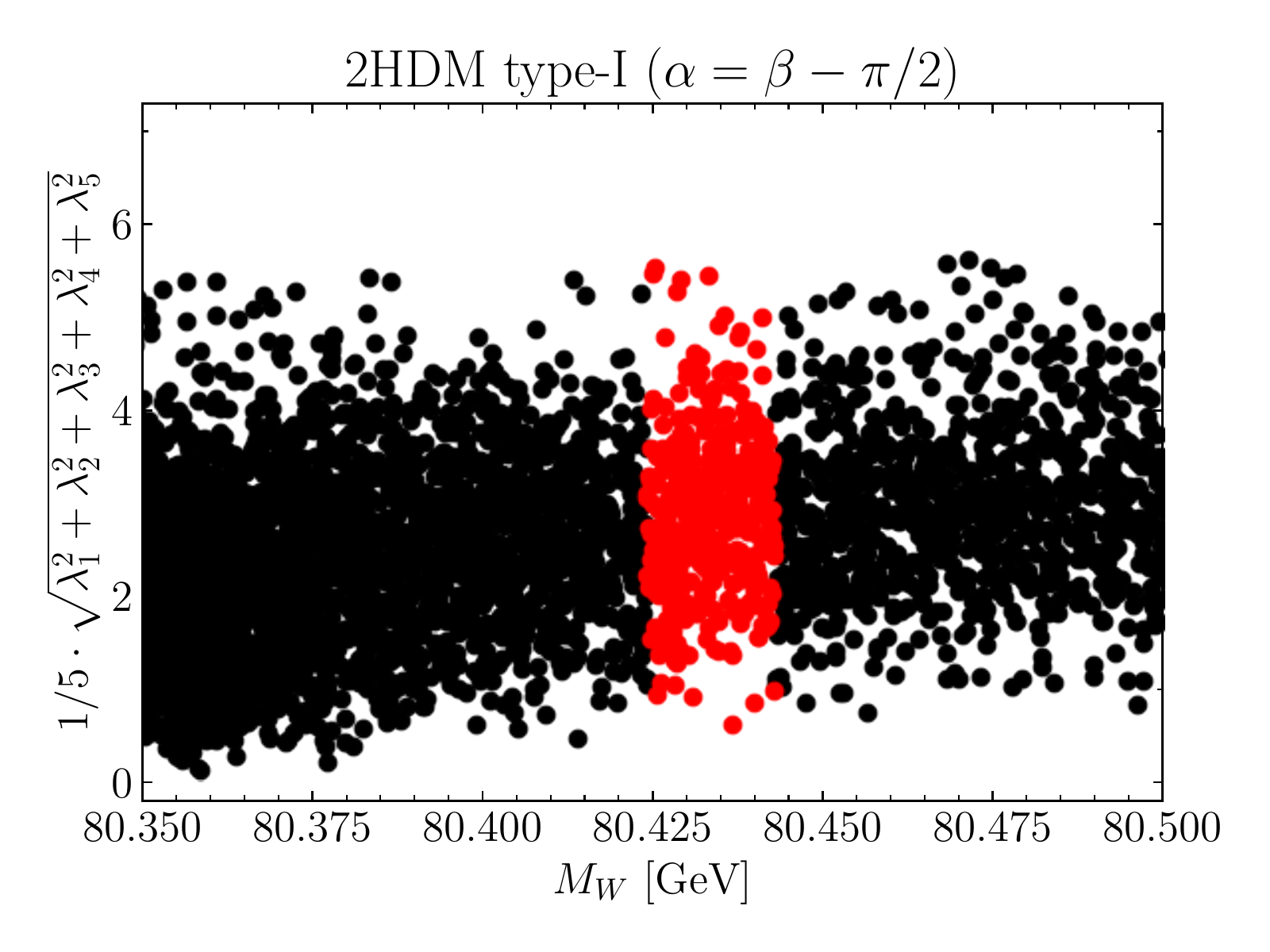} 
    \includegraphics[width=.45\textwidth]{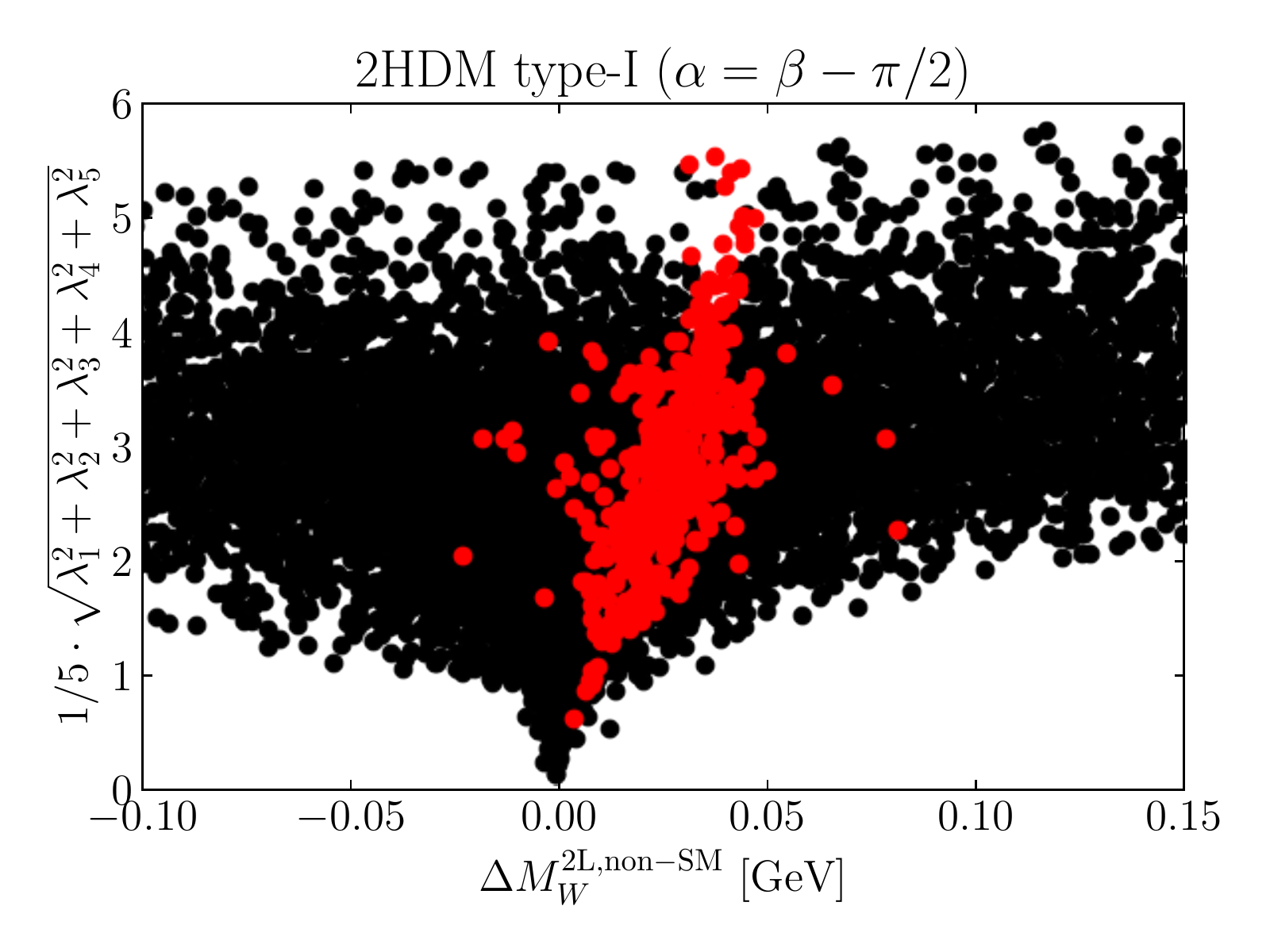}
    \caption{Parameter scan of the type-I 2HDM, where the red points are located within the $1\,\sigma$ interval of the recent \MW measurement by the CDF collaboration. In the left plot, we show for the points from the parameter scan the geometric average of the scalar quartic couplings, defined as $1/5\sqrt{\lambda_1^2+\lambda_2^2+\lambda_3^2+\lambda_4^2+\lambda_5^2}$ against the prediction for the $W$-boson mass at the two-loop level. In the right plot, we show for the same points the average of the scalar quartic couplings against the size of the two-loop non-SM corrections to $M_W$.}
    \label{fig:lam_avg}
\end{figure*}

As discussed above, sizeable differences between the BSM masses are needed to generate a significant shift of $M_W$ with respect to the SM case. Since all BSM masses can be written in the form $M^2 + \tilde\lambda v^2$ (where $\tilde\lambda$ denotes a combination of the quartic couplings and $\tan\beta$), the size of the shift to $M_W$ is correlated with the size of the quartic couplings. We show this correlation in the left panel of Fig.~(\ref{fig:lam_avg}) displaying the geometric mean of the quartic couplings in dependence of the predicted $M_W$ value. We note that all shown points fulfil the NLO unitarity constraints as explained above. As expected, higher values of $M_W$ require on average higher values of $\lambda_{1..5}$. For $M_W$ within in the $1\,\sigma$ interval of the CDF measurement, the average size of $\lambda_{1..5}$ lies between $\sim 1-5$.
It should be noted in this context that average values of $\lambda_{1..5}$ above $5$ also occur for $M_W$ values that are close to the SM value.

In the right panel of Fig.~(\ref{fig:lam_avg}) we present the same geometric average of the scalar quartic couplings as a function of the two-loop non-SM contributions to $M_W$. While it is clear that a correlation exists between the average size of the scalar quartic couplings and the magnitude (in absolute value) of the two-loop BSM contributions to $M_W$, this correlation is not very pronounced.
For instance, two-loop effects in $M_W$ of $40-50$ MeV are possible for parameter points with an average size of the quartic couplings of $\sim 2-3$. This means that significant two-loop corrections can occur even without very large couplings. 

While Fig.~(\ref{fig:lam_avg}) demonstrates that large couplings are not mandatory to obtain significant BSM contributions to $M_W$, as an additional check we have verified the behaviour under renormalisation-group (RG) running of the two example points provided in Tab.~(\ref{tab:example_points}) 
for which we generated
the necessary two-loop renormalisation-group equations with \texttt{SARAH}~\cite{Staub:2009bi,Staub:2010jh,Staub:2012pb,Staub:2013tta}. We recall that both points exhibit significant two-loop BSM effects in $M_W$. We find that the first parameter point does not encounter any Landau pole under RG running (up to the Planck scale), while for the second parameter point, Landau poles appear in the running of the scalar quartic couplings around 25 TeV well above the masses of the BSM scalars (which all have masses below 800 GeV).
Thus, the discussion of potentially large effects 
on $M_W$ within the 2HDM as an effective low-scale model with the considered mass range is not affected 
by the occurrence of nearby Landau poles.

\begin{table*}\centering
    \begin{tabular}{ccccccccc}
        \hline
        $m_H$ & $m_A$ & $m_{H^\pm}$ & $\tan\beta$ & $M^2$     & $\MW$ $[$GeV$]$ & $\MW$ $[$GeV$]$ & $\sltwo$ & $\Gamma_Z$ \\ 
        $[$GeV$]$ & $[$GeV$]$ & $[$GeV$]$       & $-$         & $[$GeV$^2]$ & (non-SM@1L) & (non-SM@2L) & $-$      & $[$GeV$]$      \\
        \hline
        853.813 & 928.352 & 809.047 & 1.206 & $444.166\times 10^3$ & 80.4001 & 80.4337 & 0.23113 & 2.4981 \\
        351.962 & 751.498 & 762.911 & 1.255 & $ 55.451\times 10^3$ & 80.3990 & 80.4339 & 0.23109 & 2.4979 \\
        \hline
    \end{tabular}
    \caption{Parameter values and results for the electroweak precision observables (calculated including two-loop non-SM corrections if not stated otherwise) for two exemplary points with \MW close to the value measured by the CDF collaboration.}
    \label{tab:example_points}
\end{table*}

\begin{figure}
    \centering
    \includegraphics[width=1.\columnwidth]{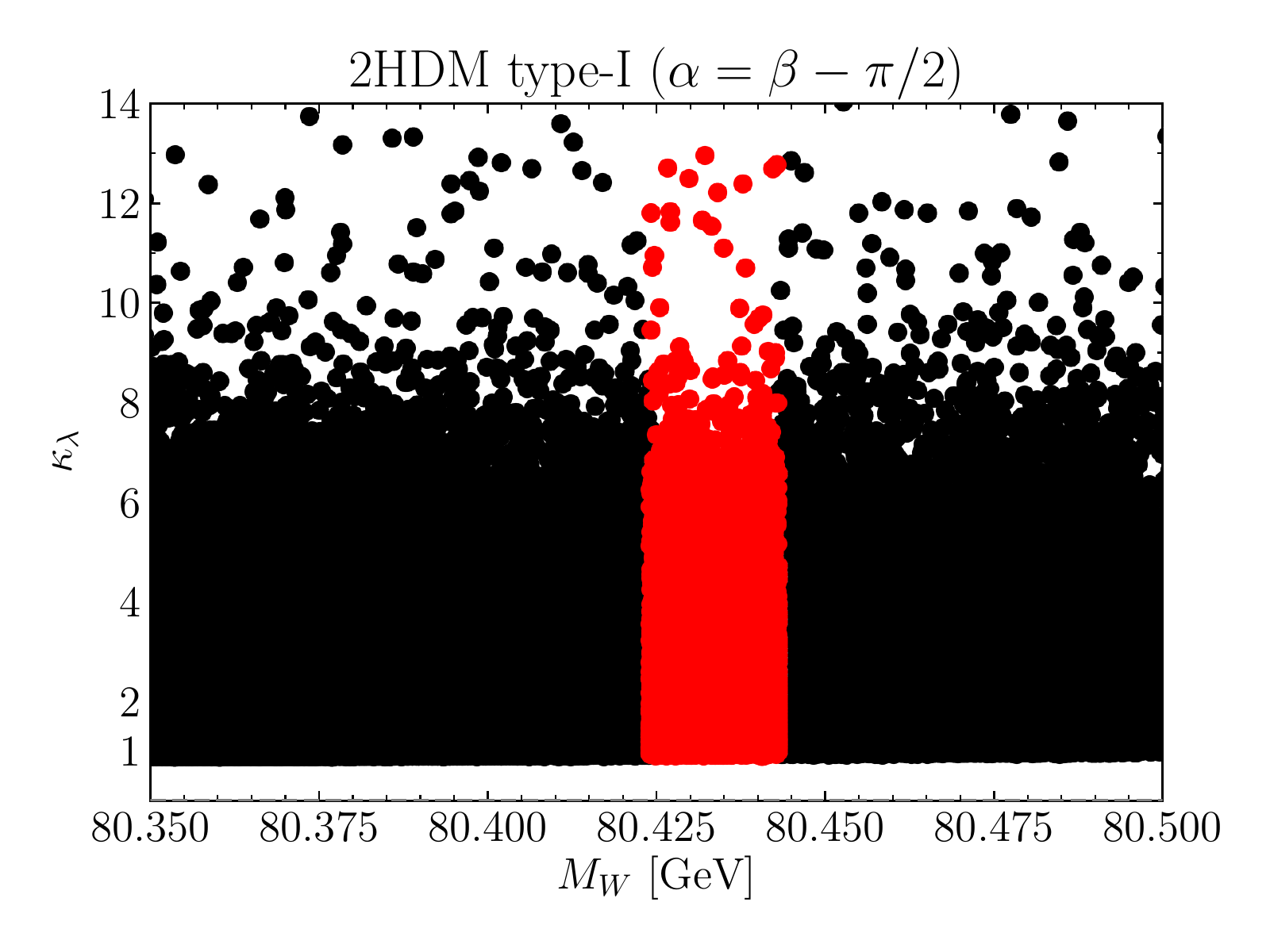}
    \caption{Parameter scan of the type-I 2HDM, where the red points are located within the $1\,\sigma$ interval of the recent \MW measurement by the CDF collaboration. We show the points from the parameter scan in the $(M_W, \kappa_\lambda)$ plane, where both quantities have been computed taking into account corrections up to the two-loop level.}
    \label{fig:MW_kappalam}
\end{figure}

Another important quantity regarding the phenomenology of the 2HDM is the trilinear Higgs coupling. As recently emphasised in Ref.~\cite{Bahl:2022jnx}, also in this case it is crucial to include higher-order corrections in order to evaluate the impact of the experimental results on the parameter space of the model. In Fig.~(\ref{fig:MW_kappalam}), we check whether the parameter region with an \MW value close to the value measured by CDF can be constrained by the  experimental limits on the trilinear Higgs coupling. We show the scan results in the $(\MW, \kappa_\lambda)$ parameter plane, where $\kappa_\lambda \equiv \lambda_{hhh}/(\lambda_{hhh}^\text{SM})^{(0)}$ is evaluated taking into account leading one-loop as well as two-loop corrections~\cite{Braathen:2019pxr,Braathen:2019zoh,Braathen:2020vwo}. We find that the results for $\kappa_\lambda$ obtained in Ref.~\cite{Bahl:2022jnx} are largely independent of the preferred region for \MW. Moreover, only a small fraction of the points within the $1\,\sigma$ interval of the CDF \MW measurement are excluded by the experimental constraint that $-1.0<\kappa_\lambda<6.6$~\cite{ATLAS:2021tyg}.


\section{Conclusions}

The mass of the $W$~boson is one of the (pseudo-) observables for which the comparison between its high-precision measurement and accurate theoretical predictions provides the highest sensitivity for discriminating between the SM and possible alternatives or extensions of it. The new measurement released by the CDF collaboration clearly has an important impact in this context. While the establishment of a new world average for \MW\ will require a careful assessment of the systematic uncertainties of the individual measurements, one can certainly expect that the incorporation of the new result announced by CDF will significantly strengthen the preference for a non-zero BSM contribution to \MW. 
Indeed, a large deviation in the $W$-boson mass
--- if confirmed by other experiments --- could 
be a hallmark of BSM physics.

In this paper, we have analysed the question of whether a deviation in \MW\ from the SM prediction as large as the one reported by CDF could actually be accommodated by well-motivated BSM models without spoiling the compatibility with the existing experimental results for other observables from different sectors and with theoretical constraints. Specifically, we have assessed the possibility that such a deviation is due to the effects of an extended Higgs sector. We have focused on the 2HDM as one of the most widely studied extensions of the SM, which can be viewed as a representative case of more complicated Higgs sectors, and we have taken into account other relevant theoretical and experimental constraints. This has led to the remarkable result that BSM quantum corrections can indeed be sufficiently large to obtain a prediction for \MW\ that is in very good agreement with the CDF measurement while being compatible with other constraints. In particular, we have demonstrated in this context the compatibility with the experimental measurements of the effective weak mixing angle and the total width of the $Z$ boson.

We have found that the mass hierarchies $m_H, m_A < m_{H^\pm}$ and $m_{H^\pm} < m_H, m_A$ are favoured by the CDF \MW measurement, whereas other mass hierarchies, and especially the case in which $m_H \sim m_A \sim m_{H^\pm}$, are disfavoured. Moreover, we have pointed out the importance of a precise evaluation of the non-SM contributions to \MW\ beyond the one-loop level. Concerning the prediction for the effective weak mixing angle at the $Z$-boson resonance, it should be noted that a sizeable quantum correction in the 2HDM bringing \MW into agreement with the CDF measurement would be favoured by the SLD measurement of the left--right asymmetry, which is interesting in view of the long-standing discrepancy between the most precise single measurements from LEP and SLC. We have also shown that the region of the 2HDM parameter space that is in agreement with the \MW value of CDF is only slightly affected by the experimental constraints on the trilinear Higgs coupling. 

We will present further details of our 2HDM results in an upcoming paper. While we focused on the 2HDM in this paper as a representative case, similar results can also be expected for other extensions of the SM Higgs sector. 


\section*{Acknowledgements}

We thank S.~Hessenberger for useful discussions and for providing us access to his code \texttt{THDM\_EWPOS}. J.B.\ and G.W. acknowledge support by the Deutsche Forschungsgemeinschaft (DFG, German Research Foundation) under Germany's Excellence Strategy -- EXC 2121 ``Quantum Universe'' – 390833306. H.B.\ acknowledges support by the Alexander von Humboldt foundation.


\bibliography{biblio}

\end{document}